\documentclass[12pt]{article}
\usepackage{amsmath,amsfonts}
\usepackage{graphicx,color,multicol,multirow}
\allowdisplaybreaks
\def\sqr#1#2{{\vcenter{\vbox{\hrule height.#2pt
\hbox{\vrule width.#2pt height#1pt \kern#1pt
\vrule width.#2pt}\hrule height.#2pt}}}}

\def\H{\widehat}
\def\h{\sqrt{h}}
\def\hN{\frac{\sqrt{h}}{N}}
\def\n{\widetilde{\nabla}}

\def\K{\widehat{K}}
\def\D{\mathcal{D}}

\def\D{\widetilde{\Delta}}
\def\N{\nonumber}
\def\be{\begin{equation}}
\def\ee{\end{equation}}
\def\beq{\begin{eqnarray}}
\def\eeq{\end{eqnarray}}

\def\sk{\smallskip\noindent}

\begin{document}

\fontfamily{cmr}\fontsize{12pt}{15pt}\selectfont
\begin{titlepage}
\null\vspace{-62pt} \pagestyle{empty}
\begin{center}
\vspace{1truein} {\Large\bfseries
Hamiltonian dynamics of gonihedric string theory} 
\vspace{6pt}
\vskip .1in
{\large Alberto Molgado$^{a,b}$ and Efra\'\i n 
Rojas$^c$}
\\
\vskip .2in
{\itshape $^a$\,Facultad de Ciencias, Universidad Aut\'onoma 
de San Luis Potos\'{\i} Campus Pedregal, Av. Parque Chapultepec
1610, Col. Privadas del Pedregal, San Luis Potos\'{\i}, 78217, M\'exico}
\\
\vskip .1in
{\itshape $^b$\,Dual CP Institute of High Energy Physics, Colima, Col, 28045,
M\'exico}
\\
\vskip .1in
{\itshape $^c$\,Facultad de F\'\i sica, Universidad Veracruzana, Cto. Gonz\'alo Aguirre Beltr\'an s/n, Zona Universitaria, Xalapa Veracruz 91000, M\'exico}
\\
\vskip .1in
\begin{tabular}{rl}
E-mail:
&{\fontfamily{cmtt}\fontsize{11pt}{15pt}\selectfont alberto.molgado@uaslp.mx}
\\
&{\fontfamily{cmtt}\fontsize{11pt}{15pt}\selectfont efrojas@uv.mx}
\end{tabular}

\fontfamily{cmr}\fontsize{11pt}{15pt}\selectfont
\vspace{.8in}

\centerline{\large\bf Abstract}
\end{center}
We develop in a consistent manner the  Ostrogradski-Hamilton 
framework for gonihedric string theory. The local action 
describing this model, being invariant under reparametrizations, 
depends on the modulus of the mean extrinsic curvature of the 
worldsheet swept out by the string, and thus we are confronted 
with  a genuine second-order in derivatives field theory. In 
our geometric approach, we consider the embedding functions as 
the field variables and, even though the highly non-linear 
dependence of the action on these variables, we are able to 
complete the classical analysis of the emerging constraints 
for which, after implementing a Dirac bracket, we are able to 
identify both the gauge transformations and the proper physical 
degrees of freedom of the model.  The Ostrogradski-Hamilton 
framework is thus considerable robust as one may recover in a 
straightforward and consistent manner some existing results 
reported in the literature. Further, in consequence of our 
geometrical treatment, we are able to unambiguously recover as 
a by-product the Hamiltonian approach for a particular relativistic 
point-particle limit associated with the gonihedric string action, 
that is, a  model linearly depending on the first Frenet-Serret 
curvature. 
\end{titlepage}

\pagestyle{plain} \setcounter{page}{2}
\fontfamily{cmr}\fontsize{12pt}{15.5pt}\selectfont

\section{Introduction}
\label{sec0}

The so-called gonihedric string model stands for a geometric 
extension for the Feynman path integral over random walks to 
an integral over random surfaces. The action functional describing 
this model is defined in a way that, when the worldsheet swept 
out by a closed string degenerates into a worldline, it must 
be reduced to the action for a point-like relativistic 
particle~\cite{Savvidy1992a,Savvidy1992b,Savvidy1993}. At the 
classical level, gonihedric string theory has also been 
considered as a model for tensionless strings associated with a 
massless spectrum of higher integer spin gauge fields~  \cite{Savvidy2002,Savvidy2003,Savvidy2003b,Savvidy2004} whereas, at the
quantum counterpart, the model describes the manner in which 
quantum fluctuations generate a nonzero string tension~\cite{Savvidy1992a,Savvidy1992b,Savvidy1993}. Further, when 
the gonihedric string model is formulated on a Euclidean lattice, 
it has been shown that it possesses a close relation to a spin 
system that generalizes the Ising model with ferromagnetic, 
antiferromagnetic and quartic 
interactions~
\cite{Savvidy1994,Savvidy1995,espriu2002,Savvidy2015,Savvidy2017}.

The underlying geometric idea behind the gonihedric action 
functional is that it is proportional to the linear size of 
a random surface and not to its area. 
Thus the action functional was originally defined as 
proportional to a ``different area element'' which was 
constructed as a combination of both the intrinsic and the 
extrinsic geometry of the worldsheet. However, in its present 
form, the action functional for the gonihedric string may be 
written in a compact and elegant form that is proportional to 
an \textit{extrinsic area element} through the modulus of the 
mean extrinsic curvature, $K^i$, where $i$ stands for the 
number of normal vectors of the worldsheet immersed into the 
background spacetime. In consequence, the action exhibits its 
second-order derivative nature through the extrinsic curvature
definition. Although the field equations resemble wave-like 
equations for a unit vector, $\K^i$, in the normal frame of the 
theory, this apparent advantage may be  deceptive since the 
$\K^i$ are non-fundamental objects which are constructed 
from the embedding functions $X^\mu$. In terms of these 
embedding functions, the field equations are fourth-order. In 
this regard, a manifestly covariant analysis on the stability 
for this class of higher-order in derivatives theories was 
recently performed~\cite{Rojas2017}. Nevertheless, to the best 
of our knowledge, a thorough Hamiltonian analysis for this 
theory has been largely overlooked, mainly due to the technical 
difficulties involved in dealing with the canonical constraint 
structure and the gauge symmetries associated with the gonihedric 
string model.

Thus, our main aim in this paper is to develop a complete 
Hamiltonian analysis of the gonihedric string theory. Our 
approach has the advantage of bringing to the forefront the 
geometrical content of the phase space associated to the model. 
In this spirit, we borrow the existent Hamiltonian formulation 
for relativistic extended objects~\cite{Rojas2004} which, in turn, 
was inspired by the Hamiltonian formulation of general relativity. 
This last formulation allows us to exploit in a great extent the 
geometry of the worldsheet in order to express the phase space 
quantities in terms of a reliable orthonormal basis. This fact 
simplifies and makes more evident the role that the Hamiltonian 
constraints play within the canonical structure. In particular, 
whenever we consider local actions describing  extended objects of 
arbitrary dimension $p$ by means of the invariants that 
characterize the geometry of the $(p+1)$-dimensional worldvolume,  
at least three essential properties are mandatory: invariance 
under diffeomorphisms of the background spacetime, invariance 
under reparametrizations of the worldvolume and, when $p+1 < N$, 
invariance under rotations of the normal vectors adapted to the 
worldvolume~\cite{Arreaga2000,Carter1996}. For the gonihedric 
string theory, by choosing the embedding functions as the 
fundamental  field variables of the theory, we are able to manage 
in a geometric manner the constraint structure of the 
Ostrogradski-Hamiltonian framework associated with the highly 
non-linear second-order in derivatives Lagrangian defining 
the model. In particular, we are capable to efficiently characterize
the constraint surface which, in turn, allows us to obtain the 
correct gauge transformations.
Indeed, for this peculiar model the last two properties give rise 
to the first- and  the second-class constraints at the Hamiltonian 
level, causing that the counting of the physical degrees of freedom 
strongly differ from what one would expect for usual extended 
objects theories~\cite{Rojas2004}. 
Also, as a by-product of our geometric formulation,  we discuss 
the point-particle limit for the gonihedric string action by 
directly comparing the higher-dimensional quantities to the one 
dimensional ones describing the relativistic curves. 

Undoubtedly, once we have the Hamiltonian framework 
at our disposal one of the next steps to follow from 
our scheme is the development and understanding of 
the naive canonical quantization of the gonihedric 
model. Unfortunately,  the highly non-linear dependence 
on the second-order derivatives of $X^\mu$ through 
the square root of $K^iK_i$, as similarly occurs in 
Born-Infeld-type actions, causes the canonical quantization 
to be rather involved. We will have occasion to explore it in 
its full generality. Regarding this, many interesting results 
have been obtained from the quantization of the model when 
considering a different set of independent field variables,  
where physical information related to the critical dimension 
and the massless spectrum of the model has been analyzed~\cite{Savvidy2003b,Savvidy2004}.

The paper is organized as follows. In Sect.~\ref{sec1}, 
we provide an overview of the gonihedric string model 
and emphasize the role that the extrinsic geometry 
plays in its description. In particular, we highligth 
the covariance of the model not only with respect 
to worldsheet diffeomorphisms but also with respect 
to local rotations of the normals to the worldsheet 
since we have a codimension of the worldsheet greater 
than one. In Sect.~\ref{sec2}, using the 
Ostrogradski-Hamilton scheme we study the canonical structure of the theory. We show that the Hamilton equations reproduce the correct equations of motion. 
In Sect.~\ref{sec3}, we develop the analysis of the constraints of the theory and make the counting of 
the physical degrees of freedom. We also analyse the infinitesimal canonical transformations associated 
to the constraints of the model. In Sect.~\ref{sec4}, 
we explore the point-particle limit for the 
gonihedric string action, emphasizing the way in 
which this limit may be achieved within our general description. We include some concluding remarks in 
Sect.~\ref{sec5} with some comments of the work as 
well as a brief discussion on the drawbacks in constructing the naive canonical quantization 
that could arise from our Hamiltonian approach. 
Finally, we collect some relevant calculus related 
to the  Poisson constraint algebra and to the 
functional derivatives of the constraints in a 
couple of appendices.

\section{The gonihedric string model}
\label{sec1}

We start with the kinematic description of the gonihedric string,
~\cite{Savvidy1992a,Savvidy1992b,Savvidy1993,Savvidy2002,Savvidy2003,
Savvidy2004,Rojas2017} $\Sigma$, described by the action 
functional\footnote{The action can be represented in the form
$
S[X^\mu] = \alpha \int_m d^2 x\,\sqrt{-g} \sqrt{ \Delta X 
\cdot \Delta X}, 
$
where $\Delta X^\mu = (1/\sqrt{-g}) \partial_a (\sqrt{-g} g^{ab} 
\partial_b X^\mu)$ (see notation below). Here, 
the second-order nature of the field theory is explicitly written. Also note, 
where appropriate hereinafter, a central dot will denote contraction with 
respect to the Minkowski metric.}
\be
S[X^\mu] = \alpha \int_m d^2x\,\sqrt{-g}\,\sqrt{K^{i}K_i},
\label{action}
\ee
where $X^\mu$ denote the embedding functions which are to be taken as the 
fundamental field variables
describing the orientable worldsheet $m$ swept out by a closed string in its
evolution in an $N$-dimensional flat Minkowski spacetime, $\mathcal{M}$, 
with metric $\eta_{\mu\nu} = \mbox{diag} (-1,1,1,\ldots, 1)$.  Lowercase Greek 
and lowercase Latin indices label the degrees of freedom in the background spacetime 
$\mathcal{M}$ and on the
worldsheet $m$, respectively, thus $\mu=0,1,\ldots,N-1$ and $a=0,1$.
In order to specify the string 
trajectory we set $x^\mu = X^\mu (x^a)$, where $x^\mu$ are 
local coordinates in the background spacetime $\mathcal{M}$ and $x^a$ are coordinates on the 
worldsheet $m$.  
Also, in action~(\ref{action}),  $K^i$ stands for the trace of 
the $i$-th extrinsic 
curvature $(i,j= 1,2,\ldots,N-2)$.  Further, the vectors $e^\mu{}_a := \partial_a X^\mu$ form a basis of 
tangent vectors to the worldsheet $m$ which allow us to define an induced metric 
on $m$ as $g_{ab} := e_a \cdot e_b$. Thus $g=\textrm{det} (g_{ab})$ stands for the determinant 
of the induced metric on the worldsheet $m$.  Finally,  $\alpha$ is a proportionality constant that 
measures the length of the surface
which corresponds to the linear size of the surface, in analogy to the definition 
of the standard Feynman path integral~\cite{Savvidy1992a,Savvidy1992b,Savvidy1993}. In 
this sense, $[\alpha] = L^{-1}$, in Planck units.

In what follows, $\nabla_a$ will denote the (torsionless) 
covariant derivative compatible with $g_{ab}$. Similarly, the 
$i$-th normal vector to $m$, $n^{\mu\,i}$,  is defined by the 
relations $e_a \cdot n^i = 0$ and $n^i \cdot n^j = \delta^{ij}$, 
where $\delta^{ij}$ is the Kronecker delta.
 It is worth noting 
that these relations determine the normal vectors $n^{\mu\,i}$  up to an $O(N-2)$ 
rotation (and a sign) and also that they  transform as a vector under normal 
frame rotations~\cite{Cheng1973,Dajczer1990}. Also note that whereas the tangential
indices are lowered and raised with $g_{ab}$ and $g^{ab}$, 
respectively, the normal indices are lowered and raised with 
$\delta_{ij}$ and $\delta^{ij}$, respectively.

The gradients of the orthonormal basis entail the definition of 
the extrinsic curvature tensor, $K_{ab}^i := - n^i \cdot \nabla_a 
\nabla_b X = K_{ba}^i$. Apart from the extrinsic curvature, 
whenever the codimension is higher than one, the extrinsic geometry of 
a surface must be complemented with the extrinsic twist potential, 
$\omega_a^{ij}$, defined by 
\be
\omega_a^{ij} := \nabla_a n^i \cdot n^j = - \omega_a^{ji}.
\label{twist}
\ee
In the case of a hypersurface embedding, the indice $i$ only takes one value 
and thus the extrinsic twist vanishes identically. Under a rotation $n^i 
\rightarrow O^i{}_j n^j$, this potential transforms as a connection 
so that, this quantity is considered as the gauge field associated 
with the normal frame rotation group~\cite{Guven1993}. Therefore, to 
implement normal frame covariance in a manifest way, we need the existence 
of a new covariant derivative $\n_a$, defined on fields transforming 
as tensors under normal frame rotations~\cite{Cheng1973,Dajczer1990,
Guven1993}, $\n_a \Phi^i{}_j := \nabla_a \Phi^i{}_j - \omega_a^{ik}
\Phi_{kj} - \omega_{a\,jk} \Phi^{ik}$. 

The variation of the action~(\ref{action}) is~\cite{Rojas2017}
\be
\label{var1}
\delta S = \alpha \int_m d^2 x\, \left( \delta (\sqrt{-g}\, ) \,\sqrt{K^iK_i}
+ \sqrt{-g}\, \delta (\sqrt{K^i K_i}\, ) \right).
\ee
The infinitesimal changes of the field variables, $X^\mu(x^a) 
\to X^\mu (x^a) + \delta X^\mu (x^a)$, may be decomposed into 
tangential and normal deformations, that is, $\delta X^\mu = 
\Phi^a e^\mu{}_a + \Phi^i n^{\mu}{}_i$ where $\Phi^a$ and $\Phi^i$ 
denote both tangential and normal deformation fields, 
respectively. On the other hand, as the action~(\ref{action}) is invariant under 
reparametrizations of the worldsheet, this gauge symmetry determines that only the 
transverse worldsheet motion is physical and, in consequence, the tangential 
deformations, $\delta_\parallel X^\mu = \Phi^a \partial_a X^\mu$, 
are usually ignored~\cite{Capovilla1995}. Therefore, it is only
necessary to consider  the normal deformations $\delta_\perp X^\mu = \Phi^i n^\mu{}_i$, 
where the $\Phi^i$ are assumed to be functions of $x^a$. 
Consequently, following the development performed 
in~\cite{Capovilla1995,Rojas2017}, the variation~(\ref{var1})
becomes
\be
\label{var3}
\delta_\perp S = - \alpha \int_m d^2 x \sqrt{-g}  \, \left( \D \K^i - 
J^{ab}_j K_{ab}^i \K^j \right) \Phi_i,
\ee
up to a total derivative term.  Here the symbol  $\D = g^{ab} \n_a \n_b$ denotes the 
worldsheet d'Alembertian operator and, $J^{ab}_i :=  g^{ab} K_i 
- K^{ab}_i$ is a tensor, and
\be
\K^i := \frac{K^i}{\sqrt{K^jK_j}},
\label{Ki}
\ee 
is a unit vector which may be thought of as defining the coordinates 
of a $S^{(N-2)}$ unit sphere, $\delta_{ij} \K^i \K^j = 1$. Evidently, 
$\K^i$ transforms like a vector under normal frame rotations. 
The tensor $J^{ab}_i$ is conserved in the sense that $\n_a 
J^{ab}_i = 0$ which stands as a consequence of the Codazzi-Mainardi 
integrability condition for surfaces immersed in a background spacetime 
is Minkowski~\cite{Kobayashi69}. Hence, the classical string trajectories 
are obtained from the $N-2$ field equations
\be
\label{eom1}
\left( \D\delta_j{}^i - J^{ab}_j K_{ab}^i 
\right)\K^j = 0.
\ee
On pedagogical grounds, this set of equations may be seen as a set 
of wave-like equations for the variables $\K^i$, but we must remember 
that the field variables are the embedding functions. 
Regarding this point, the field equations~(\ref{eom1}) are 
fourth-order in derivatives of $X^\mu$, a fact that can be directly seen
once we write the extrinsic curvature in terms of the derivatives of 
the embedding functions.
The field equations (\ref{eom1}) can also be obtained from the \textit{conserved 
stress tensor}, $f^{\mu\,a}$, associated with the model. 
Certainly, for an extended object, the conservation laws 
associated with the Poincar\'e symmetry of the ambient spacetime 
can be written in terms of a conserved tensor, $f^{\mu\,a}$, which 
turns out to be the conserved linear momentum. In this sense, 
by projecting  $\nabla_a f^{\mu\,a} = 0$ along the tangential directions 
of $m$, we obtain the gauge redundancy through Bianchi identitites 
whereas the normal projection of the conservation law, $n^i \cdot \nabla_a 
f^a = 0$, provides the equations of motion of the theory. 
Guided by the variational techniques developed in~\cite{Arreaga2000}\,, 
the conserved stress tensor is
\be
f^{\mu\,a} = - \alpha \left( \H{K}^i J^{ab}_i \,e^{\mu}{}_b + g^{ab}
\widetilde{\nabla}_b \K^i \,n^{\mu}{}_i \right).
\label{f}
\ee
From the definition and properties of the extrinsic curvature $K_{ab}^i$, 
and the Codazzi-Mainardi equation~\cite{Capovilla1995,Arreaga2000}\,, 
the field equations (\ref{eom1}) follow. Note that the stress tensor (\ref{f}) is written
in terms of the worldsheet basis, $(e^{\mu}{}_b,n^{\mu}{}_i)$. For 
Hamiltonian purposes,
a very important physical quantity can be constructed from~(\ref{f}) 
and the geometry of the string $\Sigma$ as seen from the perspective 
of the worldsheet $m$,~\cite{Rojas2004}
\be
\pi^\mu := \sqrt{h}\,\eta_a f^{\mu\,a} = - \alpha \sqrt{h}\left( 
\H{K}^i J^{ab}_i \, \eta_a e^\mu{}_b + \eta^a \widetilde{\nabla}_a 
\H{K}^i \,n^\mu{}_i \right),
\label{pi}
\ee
where $\eta^a$ is a unit time normal vector from the string $\Sigma$  
into the worldsheet $m$, and $h$ is the determinant of the spatial 
metric $h_{AB}$ to $\Sigma$, see below for details.  Note that 
uppercase Latin indices here stand for the unique value that labels the 
only spatial coordinate, usually denoted by $\sigma$, in the conventional 
string theory framework.\footnote{In fact, in the case under study here, 
the determinant $h$ corresponds to the unique component of the spatial 
metric $h_{AB}$. However, we would like to keep the notation as general 
as possible in order to easily adapt it to any generalization.}
Indeed,~(\ref{pi}) represents the conserved linear momentum density
associated with the Poincar\'e invariance of the gonihedric model 
under background translations~\cite{Arreaga2000}.

\begin{figure}[h!]
\centering
\includegraphics[width=2.9in,height=1.8in]{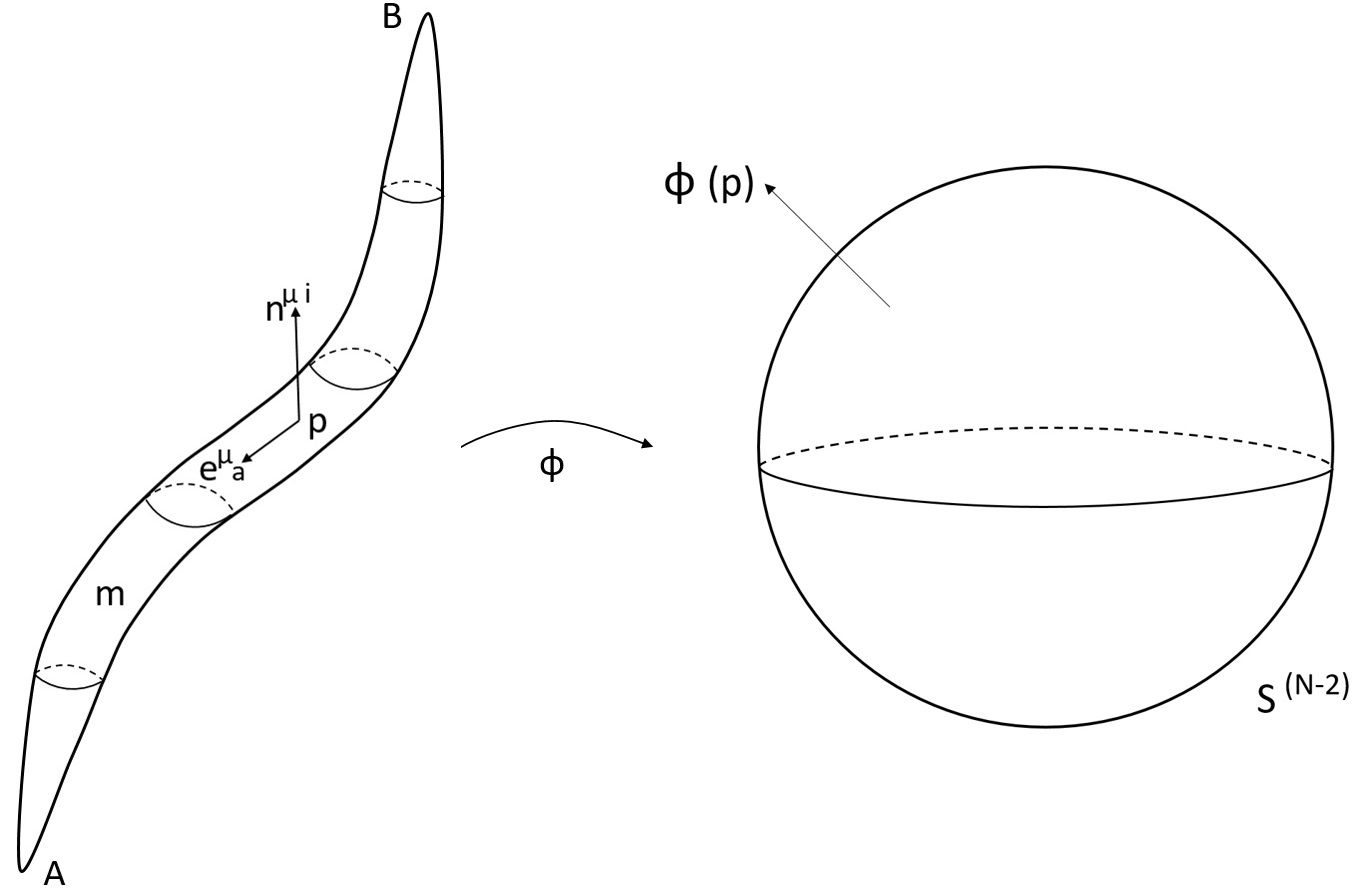}
\caption{The Gauss map $\Phi:m \rightarrow S^{(N-2)}$, maps points $p$
defined on the worldsheet to the unit sphere $S^{(N-2)}$.} 
\label{fig1}
\end{figure}

Owing to the rich geometrical nature of the model~(\ref{action}), 
the introduction of the unit vector~(\ref{Ki}) not only contributes 
to simplicity in expressing the field equations~(\ref{eom1}) but it 
also resulted convenient in order to construct the projector 
\be
\Pi ^{ij}=\delta^{ij} - \H K^i \H K^j.
\label{projector}
\ee
This projector sends fields that transform as tensors under normal frame 
rotations onto the hyperplanes transverse to the $\H{K}^i$ vectors.
Indeed, it turns out that $\Pi^{ij}\H{K}_j = 0$ and $\Pi^{ij}\n_a 
\H{K}_j = \n_a \H{K}^i$. These facts, together with the orthogonality
property $\delta_{ij}\K^i \n_a \K^j= 0$, signal that $\{ \n_a K^i,
\K^i \}$ span an orthonormal basis for a unit sphere $S^{(N-2)}$. 
Further, notice that for the case of a hypersurface, $i=1$, 
we have that $\Pi^{ij}$ vanishes. In consequence, we realize that we 
have a locally defined Gauss normal map, $\Phi :m \rightarrow S^{(N-2)}$, 
that maps points $p\in m$ to the unit sphere $S^{(N-2)}$,~Fig.~\ref{fig1}.  
Intuitively, for the present model, this Gauss map allows us to 
interpret the alleged invariance under rotations of the normal vectors 
adapted to the worldsheet, as discussed in the Introduction within the 
general extended objects setup~\cite{Kobayashi69}. 

\section{Ostrogradski-Hamilton analysis}
\label{sec2}

The general aspects of the Ostrogradski-Hamilton analysis for second-order 
in derivatives models for extended objects have been presented in detail 
in~\cite{Rojas2004}\!. Within the ADM framework adapted to extended 
objects~\cite{Rojas2000,Rojas2004}, in addition to the embedding functions, 
$X^\mu$, and their velocities, $\dot{X}^\mu$, the action~(\ref{action}) 
is also dependent on the accelerations $\ddot{X}^\mu$. On the hypersurface 
$\Sigma_t$, that is, the string $\Sigma$ considered at a fixed time $t$, 
we have the orthonormal basis $\{ \epsilon^\mu{}_\sigma, \eta^\mu, n^{\mu\,i} 
\}$, where we have defined $\epsilon^\mu{}_\sigma := \partial_\sigma X^\mu$. 
The local action of the gonihedric string field theory~(\ref{action}) can be 
written as $S[X^\mu] = \int_m d^2 x\, N\h L(g_{ab},K_{ab}^i)$ with $L = 
\alpha \sqrt{K^i K_i}$. Here, $h$ denotes the determinant of the induced 
metric on $\Sigma_t$ denoted by $h_{AB} = \eta_{\mu\nu} \epsilon^\mu{}_\sigma 
\epsilon^\nu{}_\sigma$,  where 
$\epsilon^\mu{}_\sigma$ is equivalent to the tangent vector $e^\mu{}_\sigma$ 
at any point in the hypersurface $\Sigma_t$. The mean extrinsic curvature acquires the form 
\beq
\label{eq:mean-acel}
K^i = \frac{1}{N^2} [(n^i \cdot \ddot{X}) 
- 2N^\sigma (n^i \cdot \mathcal{D}_\sigma \dot{X}) - (N^2 h^{\sigma \sigma} 
- N^\sigma N^\sigma) (n^i \cdot \mathcal{D}_\sigma \mathcal{D}_\sigma 
X)],
\eeq 
(see reference~\cite{Rojas2004} for details). The symbols $N$ and $N^\sigma$ 
denote the lapse function and the shift vector defined on $\Sigma_t$, 
respectively, obtained in a similar way as in the ADM formulation for general 
relativity~\cite{Wald}.  Note that the velocities may be written in terms 
of the lapse function and the shift vector as $\dot{X}^\mu = N \eta^\mu + 
N^\sigma \epsilon^\mu{}_\sigma$.
Also note that $\mathcal{D}_\sigma$ in~(\ref{eq:mean-acel}) stands for the 
covariant derivative such that $\mathcal{D}_\sigma h_{\sigma \sigma} = 0$.

The momenta $p_\mu$ and $P_\mu$, conjugate to $X^\mu$ and $\dot{X}^\mu$, 
respectively, in terms of the worldsheet basis, are~\cite{Rojas2004} 
\beq 
P_\mu 
& := & 
\frac{\partial L}{\partial \ddot{X}^\mu}  =  
- \frac{\h}{N} (\eta_a L^{ab}_i \eta_b) n_\mu{}^i \,,
\label{P}
\\
p_\mu 
& := &
\frac{\partial L}{\partial \dot{X}^\mu} -\frac{d\ }{d\tau}\left(\frac{\partial 
L}{\partial \ddot{X}^\mu}\right)  =  - \h \left[ \left( L\,\eta^a - 
2 \eta_b L^{c(a}_i K_c{}^{b)i} \right)e_{\mu\,a} + \eta_b \nabla_a 
\left( L^{ab}_i\,n_\mu{}^i \right) \right] 
\nonumber
\\
& & 
+\, \partial_\sigma \left[ N^\sigma P_\mu - \h (\eta_b L^{ab}_i 
\epsilon_a{}^\sigma ) n_\mu{}^i\right]\, ,
\label{p}
\eeq
where we have defined $L^{ab}_i := \partial L/\partial K_{ab}^i$. 
In our particular case we have $L^{ab}_i = \alpha \K_i g^{ab}$. 
Thus, by considering the Gauss-Weingarten equation $\n_a n^{\mu\,i} 
= K_a{}^{b\,i} \,e^{\mu}{}_b$, the momenta reduces to
\beq 
P_\mu &=& \alpha \hN \K^i\,n_{\mu\,i},
\label{Pmu}
\\
p_\mu &=& \pi_\mu + \partial_\sigma (N^\sigma P_\mu),
\label{pmu}
\eeq
where $\pi^\mu$ is defined in~(\ref{pi}) and we have considered 
the identity $g_{ab} \eta^a \epsilon^b{}_\sigma =0$. We have then 
an $8N$-dimensional phase space, $\Gamma$, spanned by $\{ X^\mu,p_\mu; 
\dot{X}^\mu,P_\mu \}$. It should be stressed that within this ADM 
approach for extended objects the boundary term $ \partial_\sigma 
(N^\sigma P_\mu )= \mathcal{L}_{\vec{N}} P_\mu$ is always 
present~\cite{Rojas2004}. In this 
extended phase space, the corresponding Legendre transformation 
$\mathcal{H}_0 := p\cdot \dot{X} + P \cdot \ddot{X} - \mathcal{L}$ with 
$\mathcal{L} = \alpha N \h \sqrt{K^i K_i}$, provides the canonical 
Hamiltonian
\be 
\mathcal{H}_0 = p \cdot \dot{X} + 2N^\sigma (P\cdot \mathcal{D}_\sigma \dot{X}) 
+ (N^2 h^{\sigma\sigma} - N^\sigma N^\sigma) (P \cdot \mathcal{D}_\sigma 
\mathcal{D}_\sigma X).
\label{H0}
\ee
We would like to emphasize the linear dependence of the canonical 
Hamiltonian on the momenta $p_\mu$ and $P_\mu$. Classically, these 
momenta can take both negative and positive values in phase space
making the canonical Hamiltonian unbounded from below, thus 
reaching both positive and negative values. 
In other words,  Ostrogradski instabilities may be present in the 
dynamics of the model. Additionally, $\mathcal{H}_0$ contains a highly 
non-linear dependence on the configuration variables $X^\mu$ and 
$\dot{X}^\mu$ by virtue of the lapse function, the shift vector and the 
induced metric.

For the model of our interest, the presence of symmetries manifests 
trough the presence of constraints. We can determine these symmetries 
by computing first the null eigenvectors of the Hessian matrix
\be 
\mathcal{H}_{\mu\nu} := \frac{\partial^2 \mathcal{L}}{\partial 
\ddot{X}^\mu \partial \ddot{X}^\nu} = \alpha \frac{\h}{N^3 
\sqrt{K^l K_l}}\Pi^{ij}n_{\mu\,i} n_{\nu\,j},
\label{hessian}
\ee
where $\Pi^{ij}$ stands for the projector defined in (\ref{projector}). One 
can easily be convinced that, due to the orthonormal property of the $\Sigma_t$ 
basis, the eigenvectors with zero eigenvalues of $\mathcal{H}_{\mu\nu}$ 
are given by $\dot{X}^\mu, \epsilon^\mu{}_\sigma$ and $P_\mu$. Guided 
by the canonical formalism developed for second-order singular 
theories~\cite{Nesterenko1989,Rojas2004}, by projecting the momenta~(\ref{Pmu}) 
along these zero-modes we get three primary constraints
\beq 
C_1 &=& P\cdot \dot{X} = 0,
\label{C1}
\\
C_\sigma &=& P \cdot \partial_\sigma X = 0,
\label{CA1}
\\
C_2 &=& N^2 P^2 - \alpha^2 h = 0.
\label{C2}
\eeq
It is convenient to turn these constraints densities into functions
in $\Gamma$. Hence, we must smear out $C_1, C_\sigma$ and $C_2$ by test
fields $\lambda, \lambda^\sigma$ and $\phi$ on $\Sigma_t$
\beq 
\mathcal{S}_\lambda &=& \int_{\Sigma_t} d\sigma \lambda\,P\cdot \dot{X},
\label{S1}
\\
\mathcal{V}_{\vec{\lambda}} &=&  \int_{\Sigma_t} d\sigma \lambda^\sigma\,
 P \cdot \partial_\sigma X,
\label{V1}
\\
\mathcal{W}_\phi &=&  \int_{\Sigma_t} d\sigma \phi\,(N^2 P^2 - \alpha^2 h).
\label{W1}
\eeq

For any two phase space functions, $F,G \in \Gamma$, the corresponding 
Poisson bracket (PB) is defined by
\be 
\{ F, G \} = \int_{\Sigma_t} d\sigma \left[ 
\frac{\delta F}{\delta X^\mu} \frac{\delta G}{\delta p_\mu} +
\frac{\delta F}{\delta \dot{X}^\mu} \frac{\delta G}{\delta P_\mu} -
\frac{\delta F}{\delta p_\mu} \frac{\delta G}{\delta X^\mu} -
\frac{\delta F}{\delta \dot{X}^\mu} \frac{\delta G}{\delta P_\mu}
\right]. 
\label{PB}
\ee
Thus, under the PB the primary constraints functions~(\ref{S1}),
(\ref{V1}) and~(\ref{W1}) are in involution 
\be
\begin{array}{ll}
\{ \mathcal{S}_\lambda , \mathcal{S}_{\lambda'} \} = 0,
& \qquad \{ \mathcal{V}_{\vec{\lambda}} , \mathcal{V}_{\vec{\lambda'}}
\} = 0,
\\
\{ \mathcal{S}_\lambda , \mathcal{V}_{\vec{\lambda}} \} =
\mathcal{V}_{\vec{\lambda'}} \qquad \lambda^{'\,\sigma} = \lambda 
\lambda^\sigma,
& \qquad \{ \mathcal{V}_{\vec{\lambda}}, \mathcal{W}_\phi \} = 0,
 \\
\{ \mathcal{S}_\lambda , \mathcal{W}_{\phi} \} = 0,
& \qquad   \{ \mathcal{W}_\phi , \mathcal{W}_{\phi'} \} = 0.
\end{array}
\label{PB1}
\ee
Therefore, we must have at least three secondary constraints. 
Following the Dirac-Bergmann procedure for constrained systems, 
this can be proved  by considering the temporal evolution of the 
primary constraints generated by the total Hamiltonian 
\be 
H_T = H_0 + \mathcal{S}_\lambda +  \mathcal{V}_{\vec{\lambda}}
+ \mathcal{W}_\phi,
\label{HT}
\ee
where $H_0 := \int_{\Sigma_t} d\sigma\,\mathcal{H}_0$. However, bearing 
in mind conditions~(\ref{PB1}), we will follow a more convenient 
strategy to obtain the secondary constraints which consists on 
projecting the momenta~(\ref{pmu}) along the null eigenvectors 
of the Hessian matrix~(\ref{hessian}), as discussed in detail 
in~\cite{Nesterenko1989}\,, and thus we 
readily obtain 
\beq 
C_3 &=&  p \cdot \dot{X} + 2N^\sigma P\cdot \mathcal{D}_\sigma \dot{X} 
+ (N^2 h^{\sigma\sigma} - N^\sigma N^\sigma) P \cdot \mathcal{D}_\sigma 
\mathcal{D}_\sigma X = 0,
\label{C3}
\\
\mathcal{C}_\sigma &=& p \cdot \partial_\sigma X + P \cdot \partial_\sigma 
\dot{X}= 0,
\label{CA2}
\\
C_4 &=& p \cdot P - P \cdot \partial_\sigma (N^\sigma P) = 0,
\label{C4}
\eeq
where we have used that $\pi \cdot P = 0$ which can be proved from 
the fact that $\K_i \n_a \K^i = 0$. As before, we can get constraint
functions in phase space by smearing out $C_3, \mathcal{C}_\sigma$ and
$C_4$ by the test fields $\Lambda, \Lambda^\sigma$ and $\Phi$ defined
on $\Sigma_t$
\beq 
S_\Lambda &=& \int_{\Sigma_t} d\sigma \,\Lambda \left[ p \cdot \dot{X} 
+ 2N^\sigma P\cdot \mathcal{D}_\sigma \dot{X} 
+ (N^2 h^{\sigma\sigma} - N^\sigma N^\sigma) P \cdot \mathcal{D}_\sigma 
\mathcal{D}_\sigma X \right],
\label{S2}
\\
V_{\vec{\Lambda}} &=& \int_{\Sigma_t} d\sigma \,\Lambda^\sigma \left( p \cdot 
\partial_\sigma X + P \cdot \partial_\sigma \dot{X} \right),
\label{V2}
\\
W_\Phi &=& \int_{\Sigma_t} d\sigma \,\phi \left[ p \cdot P - P \cdot 
\partial_\sigma (N^\sigma P)\right].
\label{W2}
\eeq
When evolving in time the secondary constraints by using the 
results $\{ S_\Lambda, H_T \} \approx 0, \{ V_{\vec{\Lambda}}, H_T \} \approx 0$ and $\{ W_\Phi, H_T \} \approx 0$ and considering the PB listed in Section~\ref{app1}, we do not obtain tertiary constraints, except for the point-particle limit to be discussed later in Section~\ref{sec4}, so the program of generating further constraints is therefore finished.

The constraints~(\ref{C1}),~(\ref{CA1}) are characteristic of second-order 
in derivatives brane models~\cite{Rojas2004}. Note that these constraints 
 just involve the 
momenta $P_\mu$, and geometrically they may be interpreted as a consequence of the orthonormality of the worldsheet basis. 
On the contrary, the constraints~(\ref{C3}) and~(\ref{CA2}) involve all the phase 
space variables. On the one hand,~(\ref{C3}) reflects the vanishing 
of the canonical Hamiltonian which is expected because of the 
invariance under reparametrizations of the theory. Certainly, it generates 
diffeomorphisms out of $\Sigma_t$ onto the worldsheet. On the other hand, 
$\mathcal{C}_\sigma = 0$, generates diffeomorphisms tangential to $\Sigma_t$.
This can be verified by considering the Poisson brackets with the 
phase space variables as we will discuss later. Regarding the
constraints~(\ref{C2}) 
and~(\ref{C4}), the first only reflects the space-like nature of 
the highest momenta $P_\mu$, whereas the latter reflects the 
orthogonality between the momenta $\pi_\mu$ and $P_\mu$. From the
geometrical point of view and regarding the Gauss map, we may depict the momenta 
$\pi_\mu$ and $P_\mu$ as vectors tangent and normal to the $S^{(N-2)}$ sphere, respectively. In fact,~(\ref{C2}) and~(\ref{C4}) will be characterized as 
second-class constraints in Dirac's terminology, as we will discuss below.   

\subsection{Hamilton equations}
\label{sec:hamilton}

In this section, we will obtain the field equations from the Hamiltonian 
point of view.  In particular, the Hamiltonian equations will be helpful 
in order to fix the Lagrange multipliers $\lambda, \lambda^A$ and 
$\phi$ in terms of the canonical variables. By using the functional
derivatives listed in Section~\ref{app0} we have first
\be
\partial_t X^\mu = \frac{\delta H_T}{\delta p_\mu} 
= \frac{\delta H_0}{\delta p_\mu}
= \dot{X}^\mu,
\ee
which is a trivial identity since the only dependence on $p_\mu$
is trough the term $p\cdot \dot{X}$ appearing in $H_0$. Second, we 
have
\beq 
\partial_t \dot{X}^\mu &=& \frac{\delta H_T}{\delta P_\mu} 
= \frac{\delta H_0}{\delta P_\mu} + \frac{\delta 
\mathcal{S}_\lambda}{\delta P_\mu} + \frac{\delta 
\mathcal{V}_{\vec{\lambda}}}{\delta P_\mu} + \frac{\delta 
\mathcal{W}_\phi}{\delta P_\mu},
\N
\\
&=& (N^2 h^{\sigma \sigma} - N^\sigma N^\sigma) \mathcal{D}_\sigma 
\mathcal{D}_\sigma X^\mu + 2 N^\sigma \mathcal{D}_\sigma \dot{X}^\mu 
+ \lambda\,\dot{X}^\mu + \lambda^\sigma \,\epsilon^\mu{}_\sigma 
\N
\\
&+& 2 \phi\, N^2 P^\mu.
\label{lagranges}
\eeq
At this stage we are able to unambiguously fix the Lagrange multipliers. Indeed, by 
contracting~(\ref{lagranges}) with the momenta $P_\mu$, and by considering 
the 
constraint~(\ref{C2}) we obtain
\be 
\phi = \frac{N}{2\alpha \h}\sqrt{K^i K_i}.
\label{lag1}
\ee
Here, we have used the general expression for the acceleration of 
an extended object (see equation~(\ref{eq:gen-accel})).
Now, contracting~(\ref{lagranges}) with $\eta^\mu$ and 
$\epsilon^\mu{}_\sigma$, respectively, and considering the constraints~(\ref{C1})
and~(\ref{CA1}) we have
\beq 
\lambda &=& \mathcal{D}_\sigma N^\sigma - \frac{N^2}{\h} \eta^a \nabla_a
\left( \frac{\h}{N} \right),
\label{lag2}
\\
\lambda^\sigma &=& N h^{\sigma\sigma} \mathcal{D}_\sigma N - N^\sigma 
\mathcal{D}_\sigma N^\sigma
+ \frac{N^2}{\h} \eta^a \nabla_a \left( \h\,\frac{N^\sigma}{N} \right),
\label{lag3}
\eeq
where we have used the time derivatives of the spatial metric $h_{\sigma\sigma}$
and its determinant (see formulae~(\ref{eq:dot-h-metric}) and~(\ref{eq:dot-h-det}), 
respectively). Next, by considering the temporal evolution of the momenta $P^\mu$, we see that the Hamilton equations
only reproduces the definition of the momenta $p_\mu$,
\beq 
\partial_t P_\mu &=& - \frac{\delta H_T}{\delta \dot{X}^\mu} 
= -\frac{\delta H_0}{\delta \dot{X}^\mu} - \frac{\delta 
\mathcal{S}_\lambda}{\delta \dot{X}^\mu} - \frac{\delta 
\mathcal{V}_{\vec{\lambda}}}{\delta \dot{X}^\mu} - \frac{\delta 
\mathcal{W}_\phi}{\delta \dot{X}^\mu},
\N
\\
&=& - p_\mu + 2 N (P \cdot \mathcal{D}_\sigma \mathcal{D}^\sigma X) \eta_\mu 
+ 2 N^\sigma (P \cdot \mathcal{D}_\sigma \mathcal{D}_\sigma X) 
\epsilon_\mu {}^\sigma -  2 (P\cdot \mathcal{D}_\sigma \dot{X}) 
\epsilon_\mu{}^\sigma  
\N
\\
&+& \mathcal{D}_\sigma (2 N^\sigma P_\mu) - \lambda \,P_\mu + 2\phi 
\,NP^2\,\eta_\mu.
\eeq
Inserting the Lagrange multipliers~(\ref{lag1}) and~(\ref{lag3}) into
this expression leads to
\beq 
p_\mu &=& \alpha \h \K_i K^i\,\eta_\mu + 2 Nh^{\sigma\sigma} (P \cdot 
\mathcal{D}_\sigma \mathcal{D}_\sigma X) \eta_\mu + 2N^\sigma 
(P \cdot \mathcal{D}_\sigma \mathcal{D}_\sigma X)\epsilon_\mu{}^\sigma
\nonumber
\\
&-& 2 (P \cdot \mathcal{D}_\sigma \dot{X})\epsilon_\mu{}^\sigma
- \alpha \h \,\eta^a \nabla_a (\K_i\, n_\mu{}^i)
+ \partial_\sigma (N^\sigma P_\mu),  
\eeq
where we have considered the fact that $\partial_t 
P_\mu = \dot{X}^a \nabla_a P_\mu = N \eta^a \nabla_a P_\mu 
+ N^\sigma \epsilon^a{}_\sigma \nabla_a P_\mu$. This expression 
matches~(\ref{p}) or~(\ref{pmu}) once we write~(\ref{pi}) in terms
of the phase space variables. Finally, the fourth set of Hamilton
equations 
\beq
\partial_t p_\mu &=& - \frac{\delta H_T}{\delta X^\mu}
= - \frac{\delta H_0}{\delta {X}^\mu} 
- \frac{\delta \mathcal{V}_{\vec{\lambda}}}{\delta {X}^\mu} 
- \frac{\delta \mathcal{W}_\phi}{\delta {X}^\mu},
\eeq
represents the field equations of motion~(\ref{eom1}) in its canonical
form as it may be straightforwardly seen by introducing the Lagrange 
multipliers~(\ref{lag1}) and~(\ref{lag2}).

\section{First- and second-class constraints}
\label{sec3}

Next, with the intention of characterizing the constraint surface, 
we will separate both the primary and secondary constraints
into first- and second-class constraints. To perform this we
construct the antisymmetric matrix composed of the PB of all the 
constraint functions\footnote{In this section, the indices $a$ 
and $b$ should not be confused with those characterizing the 
worldsheet geometry.}, $\Omega_{ab} := \{ \mathfrak{C}_a, 
\mathfrak{C}_b \}$, where we have chosen the following order
\be 
\mathfrak{C}_a = \{ \mathcal{S}_\lambda, \mathcal{V}_{\vec{\lambda}},
\mathcal{W}_\phi, S_\Lambda, V_{\vec{\Lambda}}, W_\Phi \}, \qquad
a,b= 1,2,\ldots ,6.
\ee
The matrix $\Omega_{ab}$  weakly reads
\be 
(\Omega_{ab}) \approx 
\begin{pmatrix}
0 & 0 & 0 & 0 & 0 & 0
\\
0 & 0 & 0 & 0 & 0 & 0
\\
0 & 0 & 0 & 0 & 0 & \mathbb{A}
\\
0 & 0 & 0 & 0 & 0 & \mathbb{B}
\\
0 & 0 & 0 & 0 & 0 & \mathbb{C}
\\
0 & 0 & - \mathbb{A} & \mathbb{B} & 
- \mathbb{C} & \mathbb{F}
\end{pmatrix},
\label{Omega}
\ee
where the non-vanishing entries, 
$\mathbb{A},\mathbb{B},\mathbb{C}$ and $\mathbb{F}$, are defined in Subsection~\ref{app:useful}. The rank of this matrix is 2, thus pointing 
out the existence of two second-class constraints. In order to select these second-class constraints we need to determine first the four 
zero eigenvectors $u^a_{(I)}$, such that $\Omega_{ab}u^b_{(I)} = 0$, 
with $I = 1,2,3,4$. These vectors can be taken as
\beq 
\vec{u}_{(1)} &=& (1,0,0,0,0,0),
\\
\vec{u}_{(2)} &=& (0,1,0,0,0,0),
\\
\vec{u}_{(3)} &=& (0,0,-\mathbb{B}/\mathbb{A},1,0,0),
\\
\vec{u}_{(4)} &=& (0,0,-\mathbb{C}/\mathbb{A},0,1,0).
\eeq
It is straightforward to see that the functions $\mathcal{F}_I :=
u^a_{(I)}\,\mathfrak{C}_a$ are first-class constraints. Explicitly,
\beq
\mathcal{F}_1 &=& \mathfrak{C}_1 = \mathcal{S}_\lambda,
\label{F1}
\\
\mathcal{F}_2 &=& \mathfrak{C}_2 = \mathcal{V}_{\vec{\lambda}},
\label{F2}
\\
\mathcal{F}_3 &=& \mathfrak{C}_4 - \frac{\mathbb{B}}{\mathbb{A}}
\mathfrak{C}_3 = S_\Lambda - \frac{\mathbb{B}}{\mathbb{A}} 
\mathcal{W}_\phi,
\label{F3}
\\
\mathcal{F}_4 &=& \mathfrak{C}_5 - \frac{\mathbb{C}}{\mathbb{A}}
\mathfrak{C}_3 = V_{\vec{\Lambda}} - \frac{\mathbb{C}}{\mathbb{A}} 
\mathcal{W}_\phi.
\label{F4}
\eeq
In this spirit, if we are able to choose a set of linearly independent 
vectors  $u^a_{(I')},\ (I' = 5,6)$, such that they do not depend on the 
vectors $u^a_{(I)}$ and also following the condition $\det (u^a_{(a')}) 
\neq 0$ with $a' = (I,I')$, then the functions $S_{I'} := u^a_{(I')} 
\mathfrak{C}_a$ are second-class constraints. Indeed, by choosing
\beq 
\vec{u}_5 &=& (0,0,1,0,0,0),
\\
\vec{u}_6 &=& (0,0,0,0,0,1),
\eeq
then
\beq 
\mathcal{S}_5 &=& \mathcal{W}_\phi,
\label{S11}
\\
\mathcal{S}_6 &=& W_\Phi,
\label{S22}
\eeq
are second-class constraints. Hence, the constraints $\mathcal{F}_I$ 
and $\mathcal{S}_{I'}$ define an equivalent representation of the 
constrained phase space. In this new representation of the constraint 
surface we can introduce the matrix $\mathcal{S}_{I'J'} := 
\{ \mathcal{S}_{I'}, \mathcal{S}_{J'} \}$, \,\,$I',J' = 5,6;$ and its 
inverse $(\mathcal{S}^{-1})^{I'J'}$, given by 
\be 
(\mathcal{S}_{I'J'}) = 
\begin{pmatrix}
0 & \mathbb{A}
\\
- \mathbb{A} & \mathbb{F}
\end{pmatrix}
\qquad \text{and} \qquad
\left( (\mathcal{S}^{-1})^{I'J'}\right) = \frac{1}{\mathbb{A}^2}
\begin{pmatrix}
 \mathbb{F} & -\mathbb{A}
 \\
 \mathbb{A} & 0
\end{pmatrix}.
\ee
respectively.
The inverse of the matrix $(\mathcal{S}_{I'J'})$ allows us to introduce the Dirac bracket (DB) in the usual way
\be 
\{ F, G \}_D := \{ F, G \} - \{ F, \mathcal{S}_{I'} \} (\mathcal{S}^{-1})^{I'J'}
\{ \mathcal{S}_{J'} , G \}.
\label{DB}
\ee
Once we have identified the set of second-class constraints, 
and after imposing the Dirac bracket, we can consistently set those 
constraints equal to zero~\cite{hen}\,, so that the 
first-class constraints~(\ref{F3}) and~(\ref{F4}) reduce to
\beq 
\mathcal{F}_3 &=& S_\Lambda,
\label{F32}
\\
\mathcal{F}_4 &=& V_{\vec{\Lambda}},
\label{F42}
\eeq
as expected. The counting of the physical degrees of freedom 
(dof) of the theory is as follows: 2 dof = (Total number of 
canonical variables) - 2 (number of first-class constraints) - (number of second-class 
constraints), that is,  dof $= 2 N - 5$. In the case of a gonihedric 
string immersed into a $4$-dim Minkoswki spacetime, we have dof $= 3$. 
Hence, there are three degrees of freedom for each normal vector 
of the worldsheet.

Under the Dirac bracket~(\ref{DB}), the constraint algebra reads
\be 
\begin{array}{ll}
\{ \mathcal{S}_\lambda , \mathcal{S}_{\lambda'} \}_D = 0
& \qquad
\{ \mathcal{V}_{\vec{\lambda}} , S_{\Lambda} \}_D = 
\mathcal{S}_{\mathcal{L}_{\vec{\lambda}}\Lambda} - 
\mathcal{V}_{\vec{\lambda}_1} - V_{\vec{\Lambda}_1},
\\
\{ \mathcal{S}_\lambda , \mathcal{V}_{\vec{\lambda}} \}_D = 
\mathcal{V}_{\vec{\lambda}^*}
& \qquad
\{ \mathcal{V}_{\vec{\lambda}} , V_{\vec{\Lambda}} \}_D =
\mathcal{V}_{[\vec{\lambda},\vec{\Lambda}]},
\\
\{ \mathcal{S}_\lambda , S_\Lambda \}_D = - \mathcal{S}_{\lambda_1}
- S_{\Lambda_1}
& \qquad
\{ S_\Lambda, S_{\Lambda'} \}_D = \mathcal{S}_{\lambda_2},
\\
\{ \mathcal{S}_\lambda , V_{\vec{\Lambda}} \}_D = -
\mathcal{S}_{\mathcal{L}_{\vec{\Lambda}}\lambda}
& \qquad
\{ S_\Lambda, V_{\vec{\Lambda}} \}_D = - S_{\mathcal{L}_{\vec{\Lambda}}\Lambda}
+ \mathcal{V}_{\vec{\lambda}_2},
\\
\{ \mathcal{V}_{\vec{\lambda}} , \mathcal{V}_{\vec{\lambda'}} \}_D = 0
& \qquad
\{  V_{\vec{\Lambda}}, V_{\vec{\Lambda'}} \}_D = V_{[\vec{\Lambda},
\vec{\Lambda'}]},
\end{array}
\ee
where, for the sake of simplicity, we have introduced the following definitions
\be 
\begin{array}{ll}
\lambda^{*\sigma} = \lambda \lambda^\sigma 
& \quad
\Lambda_1^\sigma = \Lambda \lambda^\sigma
\\
\lambda_1 = \lambda \Lambda
& \quad
\lambda_2 = (N^2 h^{\sigma\sigma} - N^\sigma N^\sigma)(\Lambda 
\mathcal{D}_\sigma \mathcal{D}_\sigma \Lambda' - \Lambda' 
\mathcal{D}_\sigma \mathcal{D}_\sigma \Lambda),
\\
\Lambda_1 = 2 \Lambda \mathcal{L}_{\vec{N}}\lambda
& \quad
\lambda_2^\sigma = \Lambda (N^2 h^{\sigma\sigma} - N^\sigma N^\sigma)
\mathcal{D}^2_\sigma 
\Lambda^\sigma ,
\\
\lambda_1^\sigma = 2\Lambda N^\sigma \mathcal{D}_\sigma \lambda^\sigma.
& \qquad
\end{array}
\ee
We would like to emphasize that the multiple appearance of the 
indices $\sigma$ does not indicate a tensor sum since we must 
recall that we have a unique spatial index.
It is worthwhile to emphasize that this algebra has been 
computed on the first-class constraint surface where we have taken 
the second-class constraints as strong identities after introducing 
the Dirac bracket structure~\cite{hen}.

\subsection{Infinitesimal canonical transformations}
\label{sec:canon}

Regarding our symplectic analysis, for any classical observable
$F \in \Gamma$, the Hamiltonian vector field
\be 
X_F := \int_{\Sigma_t} d\sigma \left( \frac{\delta F}{\delta p_\mu}
\frac{\delta}{\delta X^\mu} + \frac{\delta F}{\delta P_\mu}
\frac{\delta}{\delta \dot{X}^\mu} - \frac{\delta F}{\delta X^\mu}
\frac{\delta}{\delta p_\mu} - \frac{\delta F}{\delta \dot{X} \mu}
\frac{\delta}{\delta P_\mu}
\right),
\label{XF}
\ee
generates a one-parameter family of canonical transformations
$G \longrightarrow G + \delta_F G$, where $\delta_F G := \epsilon
\{ G, F \}$, with $\epsilon$ being an infinitesimal quantity. 
In particular, the Hamiltonian vector fields associated with the 
first-class constraints~(\ref{F1}),~(\ref{F2}), (\ref{F32}) 
and~(\ref{F42}), induce the infinitesimal canonical transformations
\be
\label{eq:canon-transf}
\begin{aligned}
X_{\mathcal{F}_1} \longrightarrow 
\begin{cases}
\delta_{\mathcal{S}_\lambda} X^\mu = 0,
\\
\delta_{\mathcal{S}_\lambda} \dot{X}^\mu 
= \epsilon_1\,\lambda \dot{X}^\mu, 
\\
\delta_{\mathcal{S}_\lambda} p_\mu = 0,
\\
\delta_{\mathcal{S}_\lambda} P_\mu = -\epsilon_1\,\lambda P_\mu,
\end{cases}
& \qquad \quad
X_{\mathcal{F}_3} \longrightarrow
\begin{cases}
\delta_{S_\Lambda} X^\mu = \epsilon_3\,\Lambda\dot{X}^\mu,
\\
\delta_{S_\Lambda} \dot{X}^\mu = - \epsilon_3\,\frac{\delta 
S_\Lambda}{\delta P_\mu},
\\
\delta_{S_\Lambda} p_\mu =  - \epsilon_3\,\frac{\delta 
S_\Lambda}{\delta X^\mu},
\\
\delta_{S_\Lambda} P_\mu =  - \epsilon_3\,\frac{\delta 
S_\Lambda}{\delta \dot{X}^\mu},
\end{cases}
\\
X_{\mathcal{F}_2} \longrightarrow
\begin{cases}
\delta_{\mathcal{V}_{\vec{\lambda}}} X^\mu = 0,
\\
\delta_{\mathcal{V}_{\vec{\lambda}}} \dot{X}^\mu = \epsilon_2\,
\mathcal{L}_{\vec{\lambda}} X^\mu,
\\
\delta_{\mathcal{V}_{\vec{\lambda}}} p_\mu = \epsilon_2\,
\mathcal{L}_{\vec{\lambda}} P_\mu ,
\\
\delta_{\mathcal{V}_{\vec{\lambda}}} P_\mu = 0,
\end{cases}
& \qquad \quad
X_{\mathcal{F}_4} \longrightarrow
\begin{cases}
\delta_{V_{\vec{\Lambda}}} X^\mu = \epsilon_4\,
\mathcal{L}_{\vec{\Lambda}} X^\mu,
\\
\delta_{V_{\vec{\Lambda}}} \dot{X} = \epsilon_4\,
\mathcal{L}_{\vec{\Lambda}} \dot{X}^\mu ,
\\
\delta_{V_{\vec{\Lambda}}} p_\mu = \epsilon_4\,
\mathcal{L}_{\vec{\Lambda}} p_\mu ,
\\
\delta_{V_{\vec{\Lambda}}} P_\mu = \epsilon_4\,
\mathcal{L}_{\vec{\Lambda}} P_\mu \,.
\end{cases}
\end{aligned}
\ee
Note here that the $\epsilon_i\ (i=1,\ldots,4)$ stands for 
an arbitrary gauge parameter corresponding to each of the 
first-class constraints $\mathcal{F}_i$, respectively.  
From~(\ref{eq:canon-transf}), we may infer that the constraint 
$V_{\vec{\Lambda}}$ is the generator of diffeomorphisms 
tangential to $\Sigma_t$, while $S_\Lambda$ is the generator 
of diffeomorphisms out of $\Sigma_t$ onto the worldsheet $m$. 
Analogously, $\mathcal{S}_\lambda$ is the generator of a parity 
transformation  in the sub-sector of the 
phase space given by $\{ \dot{X}^\mu; P_\mu \}$, that is,
the sector associated with the second-order in derivatives 
dependence in brane theories. Finally, we see that the vector 
constraint $\mathcal{V}_{\vec{\lambda}}$ only acts in the 
sub-sector $\{ \dot{X}^\mu; p_\mu \}$ by generating 
displacements in the orthogonal complement of this sub-sector, 
that is, in the sub-sector $\{X^\mu; P_\mu \}$.

\section{Point-particle limit of the gonihedric string action}
\label{sec4}

The kinematic geometrical description we have performed on 
the gonihedric string model is based on the Gauss-Weingarten 
equations supplemented with integrability conditions~\cite{Capovilla1995}. 
In a similar manner, there is an analogue one-dimensional description 
for this framework, used for the description of relativistic curves 
in a background spacetime~\cite{Arreaga2000}. Bearing in mind the geometric 
construction of previous sections, we are able to obtain the 
point-like analogue for the gonihedric string model\footnote{In reference~\cite{Nichols2003}\,, the author considers a different 
perspective on the Hamiltonian construction of the point-particle 
limit described here by starting from a Lagrangian in its first
order formalism.}. Indeed, by considering the orthonormal  basis $\{ \dot{X}^\mu, n^{\mu}{}_i \}$,
related to the usual Frenet-Serret basis along the particle 
trajectory $C$, we are able to establish a straightforward 
correspondence with our formulation.  This equivalence is 
compiled in Table~\ref{tabla:tab}.
\begin{table}[h!]
\centering
\begin{tabular}{ccl}
\hline
\footnotesize{Higher-dimensional quantity} & 
\footnotesize{One-dimensional analogue} & 
\footnotesize{One-dimensional description}
\\
\hline 
$g_{ab}$ & $\gamma$ & \footnotesize{Induced metric on the curve}
\\
$d^2 x \sqrt{-g}$ & $d\xi \sqrt{-\gamma}$ & \footnotesize{Element 
of line}
\\
$K_{ab}^i$ & $K^i$ & \footnotesize{Extrinsic curvature along} \\
& & 
\footnotesize{the $i$-th normal}
\\
$K^i$ & $k^i$ & \footnotesize{Mean extrinsic curvature along} \\
& & \footnotesize{the $i$-th normal}
\\
$\K^i = K^i / \sqrt{K^j K_j}$ & $\hat{k}^i = k^i/\sqrt{k^j k_j}$
& \footnotesize{Unit vector}
\\
$\omega_a{}^{ij}$ & $\omega^{ij}$ & \footnotesize{Extrinsic twist}
\\
$\n_a$ & $\n$ & \footnotesize{Covariant derivative under} \\
& & \footnotesize{rotation of the normals}
\end{tabular}
\caption{\footnotesize{Comparison among the higher-dimensional 
geometric quantities and the one-dimensional ones describing 
relativistic curves.}}
\label{tabla:tab}
\end{table}
Note that in the one-dimensional case, the overdot stands for 
the derivative with respect to an arbitrary parameter, $\xi$, 
defined on $C$. In this framework, the so-called \textit{geodesic 
curvature} $k$ given by $k=\sqrt{k^ik_i}$ is the one-dimensional
analogue of the mean extrinsic curvature for extended objects. 
Following this line of reasoning, the action functional~(\ref{action}) 
reduces to the one-dimensional field theory
\begin{equation}
\label{kaction}
S[X^\mu] = \beta \int_C d\xi\,\sqrt{-\gamma}\,k.
\end{equation}
Here, the $X^\mu$ are describing the worldline $C$. The 
corresponding field equations, $\D \hat{k}^i = 0$, are directly 
obtained from~(\ref{eom1}), by using the terms given in 
Table~\ref{tabla:tab}, and are in agreement with the results 
found in~\cite{Plyushchay1988,Plyushchay1989,
Plyushchay1990,Zoller1990,Nesterenko1996,Arreaga2000}\,. 
The classical motion is confined on a plane such that the 
linear action in $k$,~(\ref{kaction}), turns out to be topological. On 
physical grounds, this action describes the motion of massless 
particles where the constant $\beta$ plays the role of the 
helicity of the particles. 

The Hamiltonian framework for the action (\ref{kaction}) may 
be directly obtained from the analysis developed in the previous 
sections. However, we must have in mind that for the one-dimensional 
case we must not consider any contribution coming from the 
quantities characterized by the spatial index $\sigma$. 
In particular, for the one-dimensional case it turns out that, 
besides the primary and secondary constraints given by~(\ref{C1}), 
(\ref{C2}),~(\ref{C3}) and~(\ref{C4}), we have a tertiary 
constraint given by 
\be
C_5 := p^2 = 0.
\label{C5}
\ee
This extra constraint may be inferred from (\ref{BB}) by 
eliminating the spatial contributions. By applying consistency 
conditions one may find that there are no further constraints 
for the point-particle limit. It is likewise simple, from 
an Ostrogradski-Hamilton point of view, to see that we have 
a complete first-class constrained system for this theory.
A word of caution is in order. While the main symmetry for the 
gonihedric model is the invariance under reparametrizations, 
for the relativistic particle field theory (\ref{kaction}) 
there is, in addition, the presence of a $W$-symmetry which 
exhibits a deeper geometric structure. This last topic  has 
been extensively addressed in reference~\cite{Ramos1995}\,.

\section{Concluding remarks}
\label{sec5}

In this paper we have developed the Hamiltonian formulation
of the gonihedric string propagating in a Minkowski spacetime. 
Our phase space description includes the presence of four 
first-class constraints while two second-class constraints 
are present in the theory. We believe it is pertinent to mention that,
in Ref.~\cite{Savvidy2003b}\,, Savvidy considered the gonihedric 
string model in two classically equivalent versions:
firstly, a theory for which the independent field variables are 
the embedding functions, $X^\mu$, and, secondly, a theory in which 
both the embedding functions, $X^\mu$, and the induced worldsheet 
metric, $g_{ab}$, are the independent variables. 
In this paper, we have opted for the first version, the so-called model 
A in Savvidy's work, with the intention to keep the natural 
geometric structures associated with the worldsheet by means 
of the original field variables, that is, the embedding functions.    
Though the action describing the theory involves a square root 
Lagrangian that may be allegedly difficult to manipulate, we were 
able to introduce an Ostrogradski-Hamiltonian 
analysis for the constraint content of the model. In particular,
bearing in mind the previous developed geometrical description for 
extended objects, we handled in an elementary manner both the 
highly non-linear and the second-order in derivative dependence 
of the model on the embedding functions.  Thus, our approach
allowed us to construct not only the canonical Hamiltonian but also 
to identify the complete set of constraints.  After decomposing the 
set of constraints into first- and second-class, we introduced the 
Dirac bracket which  enforces the second-class constraints as strong
identities, and we were able to recognize the gauge symmetries of 
the theory by studying the infinitesimal canonical transformations 
generated by the first-class constraints. Also, as a consequence of 
the constraint characterization, we obtain the correct number of 
degrees of freedom for the  gonihedric string model.  Furthermore, 
as a byproduct of our geometrical formulation, we straightforwardly 
recover the Hamiltonian formulation for the point-particle limit of 
the gonihedric string theory by establishing a comparison between 
the higher-dimensional geometric quantities and the ones describing 
relativistic curves.   

Even though a complete Hamiltonian construction for the gonihedric string theory may be analyzed by considering the Lagrangian in its first-order formalism, the analysis introduced here allowed us to take advantage of the well-established Hamiltonian description for extended objects.  This last issue resulted completely convenient in order to characterize the constrained structure of the model. We also hope that the Hamiltonian formalism described here may pave the way for the quantum counterpart for 
the gonihedric string from a canonical perspective, as discussed for several extended objects when they are specialized to a particular geometry. Referring to this, despite that the abstract canonical quantization procedure is clear, when considering the embedding functions as the independent field variables and the highly non-linear dependence on the second-order derivatives of $X^\mu$ in the gonihedric string action, 
as similarly occurs in the Born-Infeld-type actions, the passage to the quantum theory is rather involved. Perhaps, to lighten the general problems of canonical quantization and to be able to obtain relevant physical information, it would be worth investigating only normal deformations. In such a case, the canonical constraints~(\ref{C1}), (\ref{CA1}),~(\ref{C2}) and~(\ref{CA2}) do not undergo any change 
while~(\ref{C3}) and~(\ref{C4}) get huge simplification. We will have occasion to explore it in its full generality elsewhere. In this spirit, it will be interesting to try to connect the physical implications 
arising from our Hamiltonian development with the existing quantization approach which exploits a Weyl invariance present in the so-called 
model B~\cite{Savvidy2003b,Savvidy2004}. This connection is still 
a work in progress.

\section*{Acknowledgments}
ER thanks ProdeP-M\'exico, CA-UV-320: \'Algebra, Geometr\'\i a 
y Gravitaci\'on. Also, ER thanks the Departamento de F\'\i sica 
de la Escuela Superior de F\'\i sica y Matem\'aticas del 
Instituto Polit\'ecnico Nacional, M\'exico, where this work 
was developed during a sabbatical leave. AM would like to 
acknowledge financial support from Conacyt-Mexico under project 
CB-2017-283838. This work was partially supported by Sistema 
Nacional de Investigadores, M\'exico.

\section{Appendix A. Poisson algebra of the constraints}
\label{app1}

In this appendix we summarize the Poisson algebra among the 
constraint functions $\{\mathcal{S}_\lambda,
\mathcal{V}_{\vec{\lambda}},\mathcal{W}_\phi,S_\Lambda,
V_{\vec{\Lambda}},W_\Phi\}$.   We proceed in an orderly way, 
avoiding repeated information by implicitly assuming the 
anticommutativity of the Poisson-brackets.  Please note that 
the integrals denoted by $\mathbb{A},\ \mathbb{B},\ \mathbb{C}$ 
and $\mathbb{F}$ appearing below are explicitly given 
in the Subsection~\ref{app:useful}, where we have collected some useful formulas.

\sk
First, for the constraint $\mathcal{S}_\lambda$ we obtain
\beq 
\{ \mathcal{S}_\lambda, \mathcal{S}_{\lambda'} \} &=& 0,
\\
\{ \mathcal{S}_\lambda, \mathcal{V}_{\vec{\lambda}} \} &=& 
\mathcal{V}_{\vec{\lambda'}}, \quad\qquad\qquad \qquad \qquad 
\lambda^{'\,\sigma} = \lambda\lambda^\sigma,
\\
\{ \mathcal{S}_\lambda, \mathcal{W}_{\phi} \} &=& 0,
\\
\{ \mathcal{S}_\lambda, S_\Lambda \} &=& - \mathcal{S}_{\lambda'}
- S_{\Lambda'}, \qquad\qquad \qquad \lambda' = \lambda \Lambda
\qquad \Lambda' = 2 \Lambda\mathcal{L}_{\vec{N}} \lambda,
\\
\{ \mathcal{S}_\lambda, V_{\vec{\Lambda}} \} &=& - 
\mathcal{S}_{\mathcal{L}_{\vec{\Lambda}}\lambda},
\\
\{ \mathcal{S}_\lambda, W_\Phi \} &=& W_{\lambda\Phi}. 
\eeq

\sk
Second, for the constraint $\mathcal{V}_{\vec{\lambda}}$ we get
\beq 
\{ \mathcal{V}_{\vec{\lambda}}, \mathcal{V}_{\vec{\lambda'}}\} &=& 0,
\\
\{ \mathcal{V}_{\vec{\lambda}}, \mathcal{W}_\phi\} &=& 0,
\\
\{ \mathcal{V}_{\vec{\lambda}}, S_\Lambda \} &=& \mathcal{S}_{\mathcal{L}_{\vec{\lambda}}\Lambda} 
- \mathcal{V}_{\vec{\lambda'}}
- V_{\vec{\Lambda'}},  \qquad \quad\lambda^{'\,\sigma} = 2\Lambda N^\sigma
\mathcal{D}_\sigma \lambda^\sigma \quad \,\, \Lambda^{'\,\sigma} = 
\Lambda \lambda^\sigma,
\\
\{ \mathcal{V}_{\vec{\lambda}}, V_{\vec{\Lambda}}\} &=& \mathcal{V}_{[
\vec{\lambda},\vec{\Lambda}]},
\\
\{ \mathcal{V}_{\vec{\lambda}}, W_\Phi \} &=& 0.
\eeq

\sk
Third, for the constraint $\mathcal{W}_\phi$ the PB's read
\beq 
\{ \mathcal{W}_\phi, \mathcal{W}_{\phi'}\} &=& 0,
\\
\{ \mathcal{W}_\phi, S_\Lambda \} &=& \mathcal{S}_{\lambda^*}
+ \mathcal{V}_{\vec{\lambda^*}} + \mathcal{W}_{\phi^*} -
W_{\Phi^*},
\\
\{ \mathcal{W}_\phi, V_{\vec{\Lambda}} \} &=& - 
\mathcal{W}_{\mathcal{L}_{\vec{\Lambda}}\phi},
\\
\{ \mathcal{W}_\phi, W_\Phi \} &=& - \mathcal{S}_{\lambda'}
+ \mathcal{V}_{\vec{\lambda'}} + \mathcal{W}_{\phi'}
- S_{\Lambda'} + V_{\vec{\Lambda'}} + \mathbb{A}, 
\eeq
where
\beq 
\lambda^* &=& 4\phi \Lambda N^2 h^{\sigma\sigma} (P \cdot \mathcal{D}_\sigma
\mathcal{D}_\sigma X),
\\
\lambda^{*\,\sigma} &=& -4\phi\Lambda N^2 h^{\sigma\sigma} 
(P \cdot \mathcal{D}_\sigma \dot{X}) 
\\
\phi^* &=& \phi\,\mathcal{L}_{\vec{N}}\Lambda - \Lambda\,
\mathcal{L}_{\vec{N}}\phi + 2\phi \Lambda \,Nk,
\\
\Phi^* &=& 2\phi \Lambda\,N^2,
\\
\lambda' &=& 2\,\mathcal{L}_{\vec{N}} (\phi \,\Phi\,P^2),
\\
\lambda^{'\,\sigma} &=& 2\Phi (\mathcal{D}_\sigma \phi) \,N^2P^2 
h^{\sigma\sigma} - 2\phi (\mathcal{D}_\sigma \Phi) \,(N^2 h^{\sigma\sigma} 
- N^\sigma N^\sigma)P^2 
\N
\\
&+& 2 \Phi\,N^\sigma \mathcal{D}_\sigma (\phi\,P^2 N^\sigma) - 
2\phi\Phi\,N^2h^{\sigma\sigma} (P \cdot \mathcal{D}_\sigma P),
\\
\phi' &=& 2\phi\Phi\,h^{\sigma\sigma}(P \cdot \mathcal{D}_\sigma 
\mathcal{D}_\sigma X),
\\
\Lambda' &=& 2\phi\Phi\,P^2,
\\
\Lambda^{'\,\sigma} &=& 2\phi \Phi \,N^\sigma P^2.
\eeq

\sk
Fourth, for the constraint $S_\Lambda$ we have
\beq 
\{ S_\Lambda, S_{\Lambda'} \} &=& \mathcal{S}_{\lambda'} 
\qquad \quad 
\lambda' = (N^2 h^{\sigma\sigma} - N^\sigma N^\sigma) (\Lambda 
\mathcal{D}_\sigma \mathcal{D}_\sigma \Lambda' - \Lambda' 
\mathcal{D}_\sigma \mathcal{D}_\sigma \Lambda)
\\
\{ S_\Lambda, V_{\vec{\Lambda}} \} &=& - S_{\mathcal{L}_{\vec{\Lambda}}
\Lambda} + \mathcal{V}_{\vec{\lambda'}}, 
\qquad 
\lambda^{'\,\sigma} = \Lambda (N^2 h^{\sigma\sigma} - N^\sigma N^\sigma) 
\mathcal{D}_\sigma \mathcal{D}_\sigma \Lambda^\sigma 
\N
\\
\{ S_\Lambda, W_\Phi \} &=&  \mathcal{S}_{\lambda''} + 
\mathcal{V}_{\vec{\lambda^{''}}} + \mathcal{W}_{\phi''}
- S_{\Lambda''} + V_{\vec{\Lambda^{''}}} + W_{\Phi''} + \mathbb{B},
\eeq
where
\beq 
\lambda'' &=&
- 2 \Lambda (\mathcal{L}_{\vec{N}} \Phi)h^{\sigma\sigma} 
(P \cdot \mathcal{D}_\sigma \mathcal{D}_\sigma X)
- 2\Phi \mathcal{D}_\sigma \left[ \Lambda h^{\sigma\sigma} 
(P \cdot \mathcal{D}_\sigma \dot{X}) \right],
\\
\lambda^{''\,\sigma} &=& 2\Phi  h^{\sigma\sigma}\mathcal{D}_\sigma \left[ \Lambda \left( 
N^2 h^{\sigma\sigma} - N^\sigma N^\sigma 
\right)(P \cdot \mathcal{D}_\sigma \mathcal{D}_\sigma X)\right]
\N
\\
&+& 2 \Phi  h^{\sigma\sigma} \mathcal{D}_\sigma \left[ \Lambda  N^\sigma 
(P \cdot \mathcal{D}_\sigma \dot{X}) \right] 
+ 2 \Lambda (\mathcal{L}_{\vec{N}} \Phi) h^{\sigma\sigma} 
(P \cdot \mathcal{D}_\sigma \dot{X}),
\\
\phi'' &=& \Lambda \left( h^{\sigma\sigma} - \frac{N^\sigma N^\sigma}{N^2} 
\right) \mathcal{D}_\sigma \mathcal{D}_\sigma \Phi,
\\
\Lambda'' &=& 2\Lambda\Phi h^{\sigma} (P\cdot \mathcal{D}_\sigma 
\mathcal{D}_\sigma X),
\\
\Lambda^{''\,\sigma} &=& 2\Lambda \Phi 
h^{\sigma\sigma} (P \cdot \mathcal{D}_\sigma \dot{X}),
\\
\Phi'' &=& \Lambda \mathcal{L}_{\vec{N}}\Phi - 2 \Phi \mathcal{D}_\sigma
(\Lambda N^\sigma).
\eeq

\sk
Fifth, for the constraint $V_{\vec{\Lambda}}$ we obtain
\beq 
\{ V_{\vec{\Lambda}}, V_{\vec{\Lambda'}}\} &=& V_{[\vec{\Lambda},
\vec{\Lambda'}]},
\\
\{ V_{\vec{\Lambda}}, W_\Phi \} &=& W_{\mathcal{L}_{\vec{\Lambda}}\Phi}
+ \mathbb{C}.
\eeq

\sk
Sixth, for the constraint $W_\Phi$ we only need to calculate the PB
\be 
\{ W_\Phi, W_{\Phi'} \} = \mathcal{S}_{\lambda'} + \mathcal{V}_{\vec{\lambda'}}
- V_{\vec{\Lambda'}} + \mathbb{F},
\ee
where,
\beq
\lambda' &=& h^{\sigma\sigma}\left[ \Phi \mathcal{D}_\sigma \left(P\cdot 
\mathcal{D}_\sigma (\Phi'\,P)\right) - \Phi' \mathcal{D}_\sigma 
\left(P\cdot \mathcal{D}_\sigma (\Phi\,P)\right)\right],
\\
\lambda^{'\,\sigma} &=& h^{\sigma\sigma} \left\lbrace
2\Phi' \mathcal{D}^\sigma [P \cdot \mathcal{D}_{(\sigma} (\Phi\,P) N_{\sigma)}]
+ (\mathcal{L}_{\vec{N}}\Phi') [P \cdot \mathcal{D}_\sigma (\Phi\,P)]
\right.
\N
\\
&-& \left. 2\Phi \mathcal{D}^\sigma [P \cdot \mathcal{D}_{(\sigma} 
(\Phi'\,P) N_{\sigma)}] - (\mathcal{L}_{\vec{N}}\Phi) [P \cdot 
\mathcal{D}_\sigma (\Phi'\,P)]
\right\rbrace,
\\
\Lambda^{'\,\sigma} &=& h^{\sigma\sigma} (\Phi \partial_\sigma \Phi' - 
\Phi' \partial_\sigma \Phi) \,P^2.
\eeq

\subsection{Useful formulae}
\label{app:useful}

In this subsection, we include some useful formulas that include relevant information on either the PB's described above or the Hamiltonian field equations of Section~\ref{sec:hamilton}. 

\sk
In the first instance, we consider the integrals in the previous subsection which are  explicitly given by
\beq
\mathbb{A} &=& \int_{\Sigma_t} 4\phi\Phi \,\alpha_p^2 h \,h^{\sigma\sigma} 
(P \cdot \mathcal{D}_\sigma \mathcal{D}_\sigma X),
\label{AA}
\\
\mathbb{B} &=& \int_{\Sigma_t} \Lambda \Phi \left( p^2 - 2 N^\sigma p \cdot 
\mathcal{D}_\sigma P\right) + \alpha_p^2 h \left( h^{\sigma\sigma} 
- \frac{N^\sigma N^\sigma}{N^2}
\right) \Lambda \,\mathcal{D}_\sigma \mathcal{D}_\sigma \Phi
\N
\\
&+& 4\Lambda \Phi (N^2 h^{\sigma\sigma} - N^\sigma N^\sigma) h^{\sigma\sigma} 
(P \cdot \mathcal{D}_\sigma \mathcal{D}_\sigma X) (P \cdot \mathcal{D}_\sigma
\mathcal{D}_\sigma X) 
\N
\\
&+& 8 \Lambda \Phi\, h^{\sigma\sigma} N^\sigma (P \cdot \mathcal{D}_\sigma 
\mathcal{D}_\sigma X) (P \cdot \mathcal{D}_\sigma \dot{X}) 
\N
\\
&-& 4 \Lambda \Phi h^{\sigma\sigma} (P \cdot \mathcal{D}_\sigma \dot{X})
(P \cdot \mathcal{D}_\sigma \dot{X})
+  \Lambda \Phi \,(N^2 h^{\sigma\sigma} - N^\sigma N^\sigma) (P \cdot 
\mathcal{D}_\sigma \mathcal{D}_\sigma P)
\N
\\
&+& 2 \Lambda \, (N^2 h^{\sigma\sigma} - N^\sigma N^\sigma)\,\mathcal{D}_\sigma 
\Phi \, (P \cdot \mathcal{D}_\sigma P) + N^2 h^{\sigma\sigma} \,\mathcal{D}_\sigma 
\Lambda \,\left( P \cdot \mathcal{D}_\sigma (\Phi\,P)\right)
\N
\\
&-& \Lambda\,N^\sigma \mathcal{D}_\sigma N^\sigma \,\left( P \cdot 
\mathcal{D}_\sigma (\Phi\,P)\right) + \Lambda N h^{\sigma\sigma} 
\mathcal{D}_\sigma N \,\left( P 
\cdot \mathcal{D}_\sigma (\Phi\,P)\right)
\N
\\
&+& N^\sigma N^\sigma \mathcal{D}_\sigma \Lambda \, \left( P \cdot 
\mathcal{D}_\sigma (\Phi\,P)\right) + \Lambda (\mathcal{L}_{\vec{N}}\Phi )
\left( P \cdot \mathcal{D}_\sigma (N^\sigma \,P)\right)
\N
\\
&-& 2 \Phi\, \mathcal{D}_\sigma (\Lambda N^\sigma) \left( P \cdot 
\mathcal{D}_\sigma (N^\sigma \,P)\right) - 2 \Lambda (\mathcal{L}_{\vec{N}}\Phi)
\,N^\sigma (P \cdot \mathcal{D}_\sigma P)
\N
\\
&-& 2 (\mathcal{L}_{\vec{N}}\Phi )\,\mathcal{D}_\sigma (\Lambda\,N^\sigma)
\,P^2,
\label{BB}
\\
\mathbb{C} &=& \int_{\Sigma_t} \left[  N^\sigma (\mathcal{L}_{\vec{\Lambda}}\Phi) -  
\Lambda^\sigma (\mathcal{L}_{\vec{N}}\Phi) -  \Phi 
(\mathcal{L}_{\vec{\Lambda}} N^\sigma) \right] (P\cdot \mathcal{D}_\sigma P)
\N
\\
&+& \left[  (\mathcal{D}_\sigma N^\sigma) (\mathcal{L}_{\vec{\Lambda}}\Phi) -  
(\mathcal{D}_\sigma \Lambda^\sigma) (\mathcal{L}_{\vec{N}}\Phi) -  
(\mathcal{D}_\sigma \Phi) (\mathcal{L}_{\vec{\Lambda}} N^\sigma) \right] P^2,
\label{CC}
\\
\mathbb{F} &=& \int_{\Sigma_t} 2h^{\sigma\sigma} \left( \Phi \partial_\sigma 
\Phi' - \Phi' \partial_\sigma \phi \right) \left[ P \cdot \mathcal{D}_\sigma 
\dot{X} - N^C(P\cdot \mathcal{D}_\sigma \mathcal{D}_\sigma X) \right] \,P^2.
\label{FF}
\eeq
Note that this integrals are fundamental in the definition of the Dirac bracket in 
Section~\ref{sec3}, and also to understand the emergence of the tertiary 
constraint~(\ref{C5}) in section~\ref{sec4}. 

\sk
In the second instance, in order to reproduce the Hamiltonian field equations of Subsection~\ref{sec:hamilton} we keep in mind the general expression for the acceleration of an extended object
$\Sigma$ is~\cite{Rojas2004}
\beq
\label{eq:gen-accel}
\ddot{X}^\mu &=& (\dot{N}_A + N \mathcal{D}_A N - N^B \mathcal{D}_A N_B)
\epsilon^{\mu\,A} + (\dot{N} + N^A \mathcal{D}_A N + N^A N^B k_{AB})
\eta^\mu
\N
\\
&+& (n^i \cdot \ddot{X}) n^\mu{}_i,
\eeq
together with the important formulae for the temporal derivatives of the induced 
metric and its determinant,
\beq 
\label{eq:dot-h-metric}
\dot{h}_{AB} &=& 2 N k_{AB} + 2 \mathcal{D}_{(A} N_{B)} \,,
\\
\label{eq:dot-h-det}
\dot{h} &=& 2 h ( N h^{AB}k_{AB} +  \mathcal{D}_A N^A) \,,
\eeq
respectively, where $A,B = 1,2,\ldots, p;$ here, $p$ is the 
dimension of the extended object.

\section{Appendix B. Functional derivatives of the constraints}
\label{app0}

In this Appendix, we collect the functional derivatives of the 
constraints which are relevant not only for the PB's described in Section~\ref{app1} but also for the construction of the infinitesimal 
canonical transformations of Subsection~\ref{sec:canon}.
As before, we proceed in an orderly way, by considering in the 
first step the primary constraints $\mathcal{S}_\lambda,\mathcal{V}_{\vec{\lambda}},\mathcal{W}_\phi$, while in the following steps we consider the secondary constraints $S_\Lambda,\ V_{\vec{\Lambda}}$ and $W_\Phi$, respectively.

\sk
First,
\begin{eqnarray}
\frac{\delta \mathcal{S}_\lambda}{\delta \dot{X}^\mu}
&=& \lambda\,P_\mu 
\qquad \qquad \qquad \qquad \qquad
\frac{\delta \mathcal{S}_\lambda}{\delta P_\mu}
= \lambda\,\dot{X}^\mu,
\\
\frac{\delta \mathcal{V}_{\vec{\lambda}}}{\delta X^\mu}
&=& - \partial_\sigma (\lambda^\sigma P_\mu)
= - \mathcal{L}_{\vec{\lambda}} P_\mu
\qquad \,\,
\frac{\delta \mathcal{V}_{\vec{\lambda}}}{\delta P_\mu}
= \lambda^\sigma \partial_\sigma X^\mu
=  \mathcal{L}_{\vec{\lambda}} X^\mu
\\
\frac{\delta \mathcal{W}_\phi}{\delta \dot{X}^\mu}
&=& -2\phi N P^2 \eta_\mu
\qquad \qquad \qquad \,\,
\frac{\delta \mathcal{W}_\phi}{\delta P_\mu}
= 2\phi N^2 P^\mu,
\\
\frac{\delta \mathcal{W}_\phi}{\delta X^\mu}
&=& - \partial_\sigma (2\phi N N^\sigma P^2 \,\eta_\mu)
+ \partial_\sigma (2\alpha \phi h h^{\sigma \sigma} \,\partial_\sigma 
X_\mu),
\end{eqnarray}
where $\lambda$ is a test field of weight 0 and
$\phi$ is a test field of weight -1.

\sk
Second,
\begin{eqnarray}
\frac{\delta S_\Lambda}{\delta P_\mu} &=&
\Lambda (N^2 h^{\sigma \sigma} - N^\sigma N^\sigma)
\mathcal{D}_\sigma \mathcal{D}_\sigma X^\mu
+ 2\Lambda N^\sigma \mathcal{D}_\sigma \dot{X}^\mu
\qquad \quad\,\,\,
\frac{\delta S_\Lambda}{\delta p_\mu} = \Lambda \dot{X}^\mu
\\
\frac{\delta S_\Lambda}{\delta \dot{X}^\mu} &=&
\Lambda\,p_\mu - 2 \Lambda N h^{\sigma \sigma} (P \cdot 
\mathcal{D}_\sigma \mathcal{D}_\sigma X)
\,\eta_\mu - 2\Lambda N^\sigma (P \cdot \mathcal{D}_\sigma 
\mathcal{D}_\sigma X)\,
\epsilon_\mu{}^\sigma
\N
\\
&+& 2\Lambda (P \cdot \mathcal{D}_\sigma \dot{X})\,\epsilon_\mu{}^\sigma
- \mathcal{D}_\sigma (2\Lambda N^\sigma\,P_\mu),
\\
\frac{\delta {S}_\Lambda}{\delta {X}^\mu} &=& - \mathcal{D}_\sigma
\left[ 2\Lambda N h^{\sigma\sigma} N^\sigma (P \cdot \mathcal{D}_\sigma 
\mathcal{D}_\sigma X) \,\eta_\mu \right] + \mathcal{D}^\sigma 
\left[ 2\Lambda N^\sigma (P \cdot \mathcal{D}_\sigma \mathcal{D}_\sigma 
X)\,\dot{X}_\mu \right]
\N
\\
&+& \mathcal{D}_\sigma \left[ 2\Lambda N^2 h^{\sigma\sigma} h^{\sigma\sigma} 
(P \cdot \mathcal{D}_\sigma \mathcal{D}_\sigma X)\,\epsilon_{\mu\,\sigma} 
\right] - \mathcal{D}^\sigma \left[ 2\Lambda (P \cdot \mathcal{D}_\sigma 
\dot{X}) \,\dot{X}_\mu \right]
\N
\\
&-& \mathcal{D}^\sigma \left[ 4 \Lambda N^\sigma (P \cdot \mathcal{D}_\sigma
\mathcal{D}_{\sigma} X )\,N_{\sigma} \,\epsilon_\mu{}^\sigma \right]
+ \mathcal{D}^\sigma \left[ 4\Lambda (P \cdot \mathcal{D}_{\sigma} \dot{X})
N_{\sigma} \,\epsilon_\mu{}^\sigma \right]
\N
\\
&+& \mathcal{D}_\sigma \mathcal{D}_\sigma \left[ \Lambda (N^2 h^{\sigma\sigma} 
- N^\sigma N^\sigma)\,P_\mu \right],
\end{eqnarray}
where $\Lambda$ is test field of weight 0.

\sk
Third,
\beq 
\frac{\delta V_{\vec{\Lambda}}}{\delta \dot{X}^\mu} &=&
- \partial_\sigma (\Lambda^\sigma P_\mu) 
= -\mathcal{L}_{\vec{\Lambda}} P_\mu 
\qquad \qquad
\frac{\delta V_{\vec{\Lambda}}}{\delta X^\mu} =
- \partial_\sigma (\Lambda^\sigma p_\mu) = 
-\mathcal{L}_{\vec{\Lambda}} \,p_\mu,
\\
 \frac{\delta V_{\vec{\Lambda}}}{\delta P_\mu} &=&
\Lambda^\sigma \partial_\sigma \dot{X}^\mu = 
\mathcal{L}_{\vec{\Lambda}} \dot{X}^\mu 
\qquad \qquad \qquad
 \frac{\delta V_{\vec{\Lambda}}}{\delta p_\mu} =
\Lambda^\sigma \partial_\sigma X^\mu = \mathcal{L}_{\vec{\Lambda}}
X^\mu 
\eeq

\sk
Fourth,
\begin{eqnarray}
\frac{\delta W_\Phi}{\delta X^\mu} &=& - \mathcal{D}^\sigma
\left[ P \cdot \mathcal{D}_\sigma (\Phi\,P)\,\dot{X}_\mu\right]
+ \mathcal{D}_\sigma \left[ 2 P \cdot \mathcal{D}^{\sigma} (\Phi\,P)
N^{\sigma} \,\epsilon_\mu{}_\sigma\right],
\\
\frac{\delta W_\Phi}{\delta \dot{X}^\mu} &=& P \cdot 
\mathcal{D}_\sigma (\Phi\,P)\,\epsilon_\mu{}^\sigma,
\\
\frac{\delta W_\Phi}{\delta p_\mu} &=&  \Phi \,P^\mu,
\\
\frac{\delta W_\Phi}{\delta P_\mu} &=& \Phi\,p^\mu +
(N^\sigma \mathcal{D}_\sigma \Phi - \Phi \,\mathcal{D}_\sigma N^\sigma)\,P^\mu
= \Phi \,p^\mu + \left( \mathcal{L}_{\vec{N}}\Phi \right)
\,P^\mu,
\end{eqnarray}
where $\Phi$ is a test field of weight $-1$.



\begin{thebibliography}{99}

\bibitem{Savvidy1992a} 
R. V. Ambartzumanian, G. K. Savvidy, K. G. Savvidy and G. S. 
Sukiasian, \textit{Phys. Lett.} \textbf{B275}, 99--102 (1992).

\bibitem{Savvidy1992b}
G. K. Savvidy and K. G. Savvidy, \textit{Mod. Phys. Lett.} \textbf{A8},
2963--2972 (1992), \texttt{arXiv:hep-th/9301001}.

\bibitem{Savvidy1993}
G. K. Savvidy and K. G. Savvidy, \textit{Int. J. Mod. Phys.} 
\textbf{A8}, 3993--4011 (1993), \texttt{arXiv:hep-th/9208041}.

\bibitem{Savvidy2002} 
G. K. Savvidy and R. Manvelyan, \textit{Phys. Lett.} \textbf{B533}, 
138--145 (2002), \texttt{arXiv:hep-th/0111203}.
 
\bibitem{Savvidy2003} 
A. R. Fazio and G. K. Savvidy, \textit{Mod. Phys. Lett.} \textbf{A18}, 
2817--2828 (2003), \texttt{arXiv:hep-th/0307267}. 

\bibitem{Savvidy2003b}
G. K. Savvidy, \textit{Phys. Lett.} \textbf{B552}, 72--80 (2003).

\bibitem{Savvidy2004} 
G. K. Savvidy, \textit{Int. J. Mod. Phys.} \textbf{A19}, 
3171--3194 (2004), \texttt{arXiv:hep-th/0310085}.

\bibitem{Savvidy1994}
G. K. Savvidy and F. J. Wegner, \textit{Nucl. Phys.} \textbf{B413}, 
605--613 (1994), \texttt{arXiv:hep-th/9308094}.

\bibitem{Savvidy1995}
G. K. Savvidy, K. G. Savvidy and F. J. Wegner, \textit{Nucl. Phys.} 
\textbf{B443}, 565--580 (1995), \texttt{arXiv:hep-th/9503213}.

\bibitem{espriu2002}
P. Dimopoulos, D. Espriu, E. Jan\'e and A. Prats, 
\textit{Phys. Rev. E} \textbf{66}, 056112 (2002),
\texttt{arXiv:cond-mat/0204403}

\bibitem{Savvidy2015}
G. Savvidy, \textit{Mod. Phys. Lett.} \textbf{B29}, 1550203 (2015), \texttt{arXiv:1501.01394 [hep-th]}.

\bibitem{Savvidy2017}
G. Savvidy, \textit{Gravity with linear action and gravitational 
singularities}, In Proceedings of the 17th Hellenic School and Workshops on Elementary Particle Physics and Gravity (CORFU2017), {\bf 318}, 171 (2018), 
\texttt{1705.01459 [hep-th]}.

\bibitem{Rojas2017}
E. Rojas, \textit{Int. J. Mod. Phys.} \textbf{A32}, 1750192 (2017), \texttt{arXiv:1711.01019v2 [hep-th]}.

\bibitem{Rojas2004}
R. Capovilla, J. Guven and E. Rojas, \textit{Class. Quantum Grav.} 
\textbf{21}, 5563--5585 (2004), \texttt{arXiv:hep-th/0404178}.

\bibitem{Carter1996}
B. Carter, Brane dynamics for treatment of cosmic strings and vortons, in 
\textit{``Recent Developments in Gravitation and Mathematics'', Proc. 2nd 
Mexican School on Gravitation and Mathematical Physics}, eds.  A. Garcia, 
C. Lammerzahl, A. Macias, T. Matos and D. Nu\~nez.
(Science Network Publishing, Konstanz, 1997); \texttt{arXiv:hep-th/9705172.}

\bibitem{Arreaga2000} 
G. Arreaga, R. Capovilla and J. Guven, \textit{Ann.~Phys.} 
\textbf{279}, 126--158 (2000), \texttt{arXiv:hep-th/0002088}.

\bibitem{Cheng1973}
B. Y. Cheng, \textit{Geometry of Submanifolds} (Dekker, 1973).

\bibitem{Dajczer1990}
M. Dajczer, \textit{Submanifolds and Isometric Immersions} (Publish or Perish, 1990).

\bibitem{Guven1993}
J. Guven, \textit{Phys. Rev.} \textbf{D48}, 4604--4608 (1993), \texttt{arXiv:gr-qc/9304032}.

\bibitem{Capovilla1995}
R. Capovilla and J. Guven, \textit{Phys. Rev.} \textbf{D51}, 6736--6743 
(1995), \texttt{arXiv:gr-qc/9411060v2}.

\bibitem{Kobayashi69}
S. Kobayashi and K. Nomizu, \textit{Foundations of Differential Geometry: Volume 
II} (Interscience, New York, 1969).

\bibitem{Rojas2000}
R. Capovilla, J. Guven and E. Rojas, \textit{Nucl. Phys. B Proc.~Suppl.} 
\textbf{88},337--340 (2000), \texttt{arXiv:hep-th/0004031}.

\bibitem{Wald}
R.~M.~Wald, \textit{General Relativity} (University of Chicago Press, 1984).

\bibitem{Nesterenko1989}
V. V. Nesterenko, \textit{J. Phys. A: Math. Gen.} \textbf{22}, 
1673--1687 (1989).

\bibitem{Dirac1958}
P. A. M. Dirac, \textit{Proc. Roy. Soc. A} \textbf{246} 333343 (1958)

\bibitem{hen}
M.~Henneaux and C.~Teitelboim, \textit{Quantization of Gauge Systems} (Princeton University Press, 1994).

\bibitem{Nichols2003}
A. Nichols, \textit{Acta Phys. Pol.} \textbf{B34}, 5009--5020 (2003).

\bibitem{Plyushchay1988}
M. S. Plyuschay, \textit{Mod. Phys. Lett.} \textbf{A4}, 837 (1988).

\bibitem{Plyushchay1989}
M. S. Plyuschay, \textit{Int. J. Mod. Phys.} \textbf{A4}, 
3851 (1989).

\bibitem{Plyushchay1990}
M. S. Plyuschay, \textit{Phys. Lett.} \textbf{B243}, 383 (1990).

\bibitem{Nesterenko1996}
V. V. Nesterenko, A. Feoli and G. Scarpetta, \textit{Class. Quantum 
Grav.} \textbf{13}, 1201--1211 (1996), \texttt{arXiv:hep-th/9505064}.

\bibitem{Zoller1990}
D. Zoller, \textit{Phys. Rev. Lett.} \textbf{65},  2236 (1990). 

\bibitem{Ramos1995}
E. Ramos and J. Roca, \textit{Nucl. Phys.} \textbf{B436}, 529--541 (1995), 
\texttt{arXiv:hep-th/9408019}.

\end{thebibliography}
\end{document}